\documentclass[conference]{IEEEtran}
\IEEEoverridecommandlockouts
\usepackage{cite}
\usepackage{amsmath,amssymb,amsfonts}
\usepackage{graphicx}
\usepackage{textcomp}
\usepackage{xcolor}
\usepackage{bm}
\usepackage{url}
\usepackage{subfigure}
\usepackage{algorithm}
\usepackage{algorithmicx}
\usepackage{algpseudocode}
\usepackage{cuted}

\newcommand{\pout}{p_\mathrm{out}}
\newcommand{\Prx}{P_\mathrm{Rx}}
\newcommand{\xast}{{\bm x}_\ast}

\newcommand{\prob}[1]{\mathrm{Pr}\left[{#1}\right]}
\newcommand{\erf}[1]{\mathrm{erf}\left({#1}\right)}
\newcommand{\erfinv}[1]{\mathrm{erf}^{-1}\left({#1}\right)}
\newcommand{\normal}[2]{\mathcal{N}\left({#1}, {#2}\right)}

\newcommand{\transpose}{\mathrm{T}}
\newcommand{\distance}[2]{d\left({#1},\,{#2}\right)}
\newcommand{\diag}[1]{\mathrm{diag}\left({#1}\right)}
\newcommand{\trace}[1]{\mathrm{tr}\left({#1}\right)}

\def\BibTeX{{\rm B\kern-.05em{\sc i\kern-.025em b}\kern-.08em
    T\kern-.1667em\lower.7ex\hbox{E}\kern-.125emX}}
\begin{document}

\graphicspath{{../fig/}}

\title{Mitigating the Impact of Location Uncertainty on Radio Map-Based Predictive Rate Selection\\via Noisy-Input Gaussian Process}

\author{\IEEEauthorblockN{Koya Sato}
\IEEEauthorblockA{Artificial Intelligence eXploration Research Center,\\
The University of Electro-Communications, 1-5-1, Chofugaoka, Chofu-shi, Tokyo, Japan \\
E-mail: k\_sato@ieee.org}
\thanks{This work supported by JST, PRESTO, Grant JPMJPR23P3, and JSPS KAKENHI Grant Number 25K00138.}
}

\maketitle

\begin{abstract}
    This paper proposes a predictive rate-selection framework based on Gaussian process (GP)-based radio map construction that is robust to location uncertainty. Radio maps are a promising tool for improving communication efficiency in 6G networks. Although they enable the design of location-based maximum transmission rates by exploiting statistical channel information, existing discussions often assume perfect (i.e., noiseless) location information during channel sensing. Since such information must be obtained from positioning systems such as global navigation satellite systems, it inevitably involves positioning errors; this location uncertainty can degrade the reliability of radio map-based wireless systems. To mitigate this issue, we introduce the noisy-input GP (NIGP), which treats location noise as additional output noise by applying a Taylor approximation of the function of interest. Numerical results demonstrate that the proposed NIGP-based design achieves more reliable transmission-rate selection than pure GP and yields higher throughput than path loss-based rate selection.
\end{abstract}


\section{Introduction}
\label{sect:intro}
Radio maps have attracted attention for next-generation wireless systems\cite{zeng_comst2024}.
A radio map is a tool for visualizing communication quality, providing statistical information on the wireless channel.
Although related discussions were initiated in the context of dynamic spectrum sharing\cite{zhao-dyspan2007}, recent studies have demonstrated advances in various applications such as resource allocation\cite{perez_iotj2025} and localization\cite{tian_jiot2025}.
\par
Major research viewpoints on radio maps can be classified into (i)\,the accurate construction of the radio map, or (ii)\,its integration into wireless systems.
Several works addressing (i) have revealed that an accurate radio map can be constructed through crowdsensing-based on-site measurements combined with spatial interpolation\cite{wang_twc2024}.
In this direction, a server requests mobile nodes to widely measure the received signal power values.
Once sufficient data is collected, the server can construct a radio map reflecting detailed site-specific performance regarding the radio environment by estimating the missing information based on a spatial regression for the collected data.
In particular, Gaussian process regression (GPR)\cite{Rasmussen2004} can achieve optimal predictive accuracy while simultaneously quantifying prediction uncertainty.
This feature facilitates discussions concerning (ii); for example, previous work in \cite{kallehauge_gcom2022} has proposed a GPR-based transmission rate prediction framework for ultra-reliable low latency communication applications.
This method predicts the maximum achievable transmission rate as a function of the receiver’s coordinates under a given outage probability constraint.
Storing these pre-computed rates at the transmitter can significantly reduce the overhead required for estimating communication quality between transmitters and receivers during rate calculation, thereby enabling low latency and high reliability.
Similarly, the works in \cite{sato-tccn2017}\cite{5711699} proposed transmission power optimization frameworks based on Kriging, a spatial interpolation method equivalent to GPR\cite{Cressie}, which satisfies interference-outage constraints in spectrum-sharing environments.
As these works demonstrated, GPR-based radio maps enhance the accuracy of radio propagation estimation and improve the reliability and efficiency of wireless communication systems.
\par
However, existing studies on applications of radio maps often assume perfect location information in crowdsensing.
In practice, the sensing nodes must capture the location information based on their positioning capability.
For instance, if the sensing nodes use the global navigation satellite system (GNSS), its performance significantly depends on line-of-sight availability between satellites and nodes; positioning errors ranging from several meters to tens of meters can occur depending on the environment, thereby degrading the reliability of the GPR-based communication design.
Although numerous technologies for improving positioning accuracy have been proposed\cite{Zangenehnejad2021}, there is typically a trade-off between accuracy and device cost.
Since crowdsensing systems rely on off-the-shelf mobile devices, minimizing the device cost is essential. Against this background, we aim to develop a design framework for reliable communications based on a radio map with limited-accuracy location information.
\par
In this paper, we propose a rate prediction framework based on noisy-input GP (NIGP) that is robust to location uncertainty.
NIGP is a variant of GPR explicitly accounts for the input noise\cite{NIPS2011_a8e864d0}.
Unlike the pure GP, which considers the sensing noise only, NIGP recasts the input noise as additional output noise by approximating the function of interest based on the first-order Taylor approximation, thereby improving robustness to the input noise.
Since the NIGP is based on the Taylor approximation, we discuss the following research questions:
\begin{itemize}
    \item Can NIGP enhance communication reliability in GP-based rate prediction?
    \item Can improving the approximation accuracy in NIGP further enhance the reliability of rate prediction?
\end{itemize}
To clarify these questions, we present the rate selection methods based on two NIGP-based methods: the original NIGP based on the first-order Taylor approximation and its extension to the second-order approximation.
Numerical results demonstrate that the NIGP can improve outage probability performance while maintaining high-transmission rate selection performance.
\\
\textit{Notations:}
Throughout this paper, $\mathbf{I}_{(n\times n)}$ denotes the $n\times n$ identity matrix.
The operators $\mathrm{Cov}[z_i, z_j]$, $\mathbb{E}[\cdot]$, and $\mathrm{Var}[\cdot]$ represent covariance between random variables $z_i$ and $z_j$, expectation, and variance, respectively, while $\normal{\mu}{\sigma^2}$ denotes a Gaussian distribution with mean $\mu$ and variance $\sigma^2$.
Moreover, we define $\distance{\bm{x}_i}{\bm{x}_j}$ as the Euclidean distance between the vectors $\bm{x}_i$ and $\bm{x}_j$.
Finally, $(\cdot)^\transpose$ denotes the transpose operation.
\section{System Model}
\label{sect:systemmodel}
This paper discusses predicting the transmission rate based on a sensing-based radio map construction result under the location uncertainty.
Possible applications include predictive resource allocation in cellular networks\cite{kallehauge_gcom2022} and vehicle-to-infrastructure (V2I) communications\cite{yi-tnse2024}.
\subsection{Sensing Model}
Consider an outdoor radio map construction scenario where a transmitter of interest is located at the coordinate ${\bm x}_\mathrm{Tx} \in \mathbb{R}^2$.
The main task of a radio map is to visualize the received signal power on the area of interest.
To accurately construct it, we assume wireless devices equipped with multiple-antennas (e.g., smartphones or autonomous vehicles) independently measure the received signal power values at own locations; the sensing results are then reported to a centralized server.
\par
If a device can sufficiently average the effects of mutipath fading owing to its multiple antennas, the received signal power at ${\bm x}$ can be modeled as,
\begin{align}
     \Prx({\bm x}) &= P_\mathrm{Tx} - 10\eta\log_{10}d({\bm x}_\mathrm{Tx}, {\bm x}) + W({\bm x}) \triangleq f({\bm x})
\end{align}
where $P_\mathrm{Tx}$ is the transmission power\,[dBm] and $\eta$ is the path loss index.
Further, $W({\bm x})$ represents the shadowing at ${\bm x}$, modeled as a spatially-correlated log-normal distribution with zero mean, standard deviation $\sigma_\mathrm{dB}$, and the correlation distance $d_\mathrm{cor}$\,[m].
Based on Gudmundson's empirical shadowing model\cite{Gudmundson-el1991}, the covariance function of shadowing values between two locations can be given by,
\begin{equation}
    \mathrm{Cov}\left[W({\bm x}_i), W({\bm x}_j)\right] = \sigma_\mathrm{dB}^2 \exp\left(-\frac{d({\bm x}_i, {\bm x}_j)}{d_\mathrm{cor}}\ln2\right).
    \label{eq:covariance_shadowing}
\end{equation}
\par
The radio map construction problem typically divides the two-dimensional area into squared grids to efficiently manage the data; in contrast, it is preferable to choose the grid length fully shorter than $d_\mathrm{cor}$ to construct a high-resolution radio map.
To discuss the effects of location uncertainty on a high-resolution radio map, we assume sufficiently small grid lengths.
If a sensing location is obtained by the node's positioning capability, the sensing location with noise can be derived as the following equation:
\begin{equation}
    \tilde{\bm x} = {\bm x} + {\bm \epsilon}_x,
\end{equation}
where ${\bm \epsilon}_x$ is the error for the location estimation following ${\bm \epsilon}_x \sim \normal{{\bm 0}_2}{\sigma^2_x {\bf I}_{(2\times2)}}$, where ${\bm 0}_2$ is the two-dimensional zero vector.
Consider the received signal power values are measured at $N$ locations (${\bm x}_i, i \in\{1, 2,\cdots, N\}$).
The $i$-th measurement value can be expressed as,
\begin{equation}
    y(\tilde{\bm x}_i) = f({\bm x}_i) + \epsilon_y,
\end{equation}
where $\epsilon_y \sim \normal{0}{\sigma_y^2}$ is the observation error caused by uncertainty in measurement; e.g., thermal noise, quantization error, or imperfection in averaging multipath fading.
After $N$ sensing results are reported, the sever can then construct the dataset, $\mathcal{D} = \left\{\left[\tilde{\bm x}_i, y(\tilde{\bm x}_i)\right]\mid i=1, 2, \cdots, N\right\}$.
Owing to this dataset and a regression method, we first interpolate the received signal power at a test location $\xast$ where no direct sensing information is available.
In particular, this step tries to infer the statistical performance on $\Prx(\xast)$ based on a GPR to design the transmission rate for the receiver at $\xast$.
Note that, throughout this paper, we discuss the performance on a test location $\xast$; however, by regarding each grid in the area as a test point and applying the GP at each location, we can construct a radio map for the entire area\footnote{This paper does not consider the error in the test location $\xast$. However, even if it contains the uncertainty (e.g., $\tilde{\bm x}_\ast \sim \normal{\xast}{\sigma^2_x}$), our method can work by iterating i)\,sampling $\tilde{\bm x}_\ast$, ii)\,predicting the rate on $\tilde{\bm x}_\ast$ based on the proposal in Sect.\,\ref{sect:proposed}, and iii)\,computing the empirical cumulative distribution of the predicted rate. Since it is a straightforward extension, this paper focuses on how the rate selection works on $\xast$.}.
\par
Note that, to discuss the effects of function approximation in NIGP, we assume parameters for prior distribution regarding the radio map (i.e.,\,$\sigma_\mathrm{dB}$, $d_\mathrm{cor}$, and $\sigma_x$) are available at the server as prior information; however, these parameters can be estimated by a maximum-likelihood estimation for the multivariate normal distribution\,\cite{Rasmussen2004}\cite{NIPS2011_a8e864d0}.
Further, because the objective of the radio map construction is to tune the communication performance of the transmitter at the operator side, $P_\mathrm{Tx}$ and ${\bm x}_\mathrm{Tx}$ are assumed to be also available.
\subsection{Problem Setting}
Our discussion aims to find an appropriate downlink transmission rate so that the receiver on ${\bm x}_\ast$ can decode the message as fast as possible with high reliability.
One of the simple strategies for the rate selection is to compute the channel capacity on $\xast$ based on the radio map result and treat it as the transmission rate without consideration for the estimation error in the radio map construction.
However, this strategy often contains an estimation error depending on the conditions of the dataset, such as $N$ and $\sigma_x$; i.e., rate selection without consideration for the error in the radio map construction may lead to communication failure.
Thus, we define this problem as a rate selection problem constrained by an outage probability.
\par
When the channel state on ${\bm x}$ is static, the channel capacity at ${\bm x}$ can be given by $C({\bm x}) = \log_2\left(1+\gamma({\bm x})\right)$\,[bps/Hz], 
where $\gamma({\bm x}) = 10^{(\Prx({\bm x}) - N_0)/10}$ is the signal-to-noise ratio (SNR).
When the transmission rate for $\xast$ is denoted by $R(\xast)$, a message from the transmitter with this rate can be successfully decoded if $R(\xast)\leq C(\xast)$.
Thus, introducing a target outage probability $\pout$, this problem can be written by the following:
\begin{subequations}\label{eq:optimization}
\begin{align}
    \operatorname{maximize}\quad & R({\bm x}_\ast) \label{eq:objective} \\
    \text{subject to}\quad & \mathrm{Pr}\big[R({\bm x}_\ast) \leq C({\bm x}_\ast)\big] \geq 1 - \pout \label{eq:constraint_outage}
\end{align}
\end{subequations}
By precomputing solutions to this problem for every possible point associated with $\xast$, the BS can determine a transmission rate that satisfies the outage constraint for each receiver coordinate.
Note that we do not touch on the detailed modulation format to satisfy the transmission rate; one practical solution is using adaptive modulation and coding capability.
\section{Rate Selection With Pure GP}
\label{sect:exactgp}
As a baseline, this section presents a rate selection via the pure GP without consideration for the location uncertainty (i.e., $\sigma_x=0$)\cite{Rasmussen2004}\cite{kallehauge_gcom2022}.
Let us define an $N$-dimensional input vector ${\bf X}=\left[\tilde{\bm x}_1, \tilde{\bm x}_2, \cdots, \tilde{\bm x}_N\right]^\transpose$ and an $N$-dimensional output vector ${\bf Y} = \left[y(\tilde{\bm x}_1), y(\tilde{\bm x}_2), \cdots, y(\tilde{\bm x}_N)\right]^\transpose$.
When $\sigma_x=0$, the observation vector ${\bf Y}$ follows
$    {\bf Y}_{\mid \sigma_x=0} \sim \normal{\overline{\bf P}}{{\bf K}+\sigma^2_y {\bf I}_{(N\times N)}}$,
where $\overline{\bf P}$ is the $N$-dimensional vector in which its $i$-th element represents $\overline{P}(\tilde{\bm x}_i\mid \sigma_x=0) = P_\mathrm{Tx} - 10\eta\log_{10}\distance{{\bm x}_\mathrm{Tx}}{\tilde{\bm x}_i}$, and ${\bf K}$ is the $N\times N$ covariance matrix where its $(i,j)$-th element represents the kernel function $k({\bm x}_i, {\bm x}_j)$.
When the shadowing shows the spatial correlation based on Eq.\,\eqref{eq:covariance_shadowing}, it can be modeled with the exponential kernel defined by,
\begin{equation}
    k(\tilde{\bm x}_i, \tilde{\bm x}_j) = \sigma^2_k \exp\left(-\frac{\distance{\tilde{\bm x}_i}{\tilde{\bm x}_j}}{l^2}\right),
    \label{eq:exponential_kernel}
\end{equation}
where $\sigma_k$ and $l$ are hyperparameters.
Joint distribution of the observations and the test point is given by,
\begin{equation}
    \begin{bmatrix}
      {\bf Y} \\
      y(\xast)
    \end{bmatrix}\sim
    \normal{
        \begin{bmatrix}
          \overline{\bf P} \\
          \overline{P}(\xast)
         \end{bmatrix}
    }{
         \begin{bmatrix}
            {\bf K} + \sigma^2_y {\bf I}_{(N\times N)} & {\bm k}({\bf X}, \xast) \\
            {\bm k}({\bf X}, \xast)^\mathrm{T} & k(\xast, \xast)
         \end{bmatrix}
    },
\end{equation}
where ${\bm k}({\bf X}, \xast) \triangleq \left[k(\tilde{\bm x}_1, {\xast}), k(\tilde{\bm x}_2, {\xast}), \cdots, k(\tilde{\bm x}_N, {\xast})\right]^\mathrm{T}$.
Based on this relationship, the expectation and variance for $\Prx(\xast)$ can be given by the following equations, respectively:
\begin{align}
    P_\mathrm{EG}({\bm x}_\ast) &= \mathbb{E}\left[\Prx({\bm x}_\ast)\mid \mathcal{D}, \sigma_x=0\right] \triangleq \mathrm{GP_{mean}} ({\bf K}')\nonumber\\
    &= \overline{P}(\xast) + {\bm k}({\bf X}, \xast)^\mathrm{T} \left({\bf K}'\right)^{-1}\left({\bf Y} - \overline{\bf P}\right),\label{eq:exactgp_mean}\\
    \sigma^2_\mathrm{EG}({\bm x}_\ast) &\triangleq \mathrm{Var}\left[\Prx({\bm x}_\ast)\mid \mathcal{D}, \sigma_x=0\right] \triangleq \mathrm{GP_{var}}({\bf K}')\nonumber\\
    &= k(\xast, \xast) - {\bm k}({\bf X}, \xast)^\mathrm{T} \left({\bf K}'\right)^{-1}{\bm k}({\bf X}, \xast),\label{eq:exactgp_var}
\end{align}
where ${\bf K}' \triangleq {\bf K}+\sigma^2_y {\bf I}_{(N,N)}$.
This computation updates the statistical knowledge regarding $\Prx(\xast)$ from the prior distribution $\Prx(\xast) \sim \normal{\overline{P}(\xast\mid \sigma_x=0)}{\sigma^2_\mathrm{dB}}$ to its posterior distribution $\Prx(\xast\mid \mathcal{D}) \sim \normal{P_\mathrm{EG}(\xast)}{\sigma^2_\mathrm{EG}}$, thereby improving the communication efficiency.
\par
Next, based on the posterior distribution, we solve the rate selection problem defined in Eqs.\,\eqref{eq:objective}\eqref{eq:constraint_outage}.
Since the outage probability for the rate selection monotonically increases with the increase in the transmission rate, the rate can be maximized when the outage probability satisfies
\begin{align}
    \prob{R(\xast)\leq C(\xast)} &= 1-\pout.
\end{align}
Assuming that $\Gamma(\xast)\triangleq 10\log_{10}\gamma(\xast)$ and $r(\xast)\triangleq 10\log_{10}\left(2^{R(\xast)}-1\right)$, this equation can be then converted to
\begin{align}
    \prob{\Gamma(\xast) \geq r(\xast)} &= \nonumber\\
    \frac{1}{2}\Bigg(1-&\erf{\frac{r(\xast)-P_{\mathrm{EG}}(\xast)+N_0}{\sqrt{2}\sigma_\mathrm{EG}}}\Bigg).
\end{align}
Based on this equation, by solving $\prob{\Gamma(\xast) \geq r(\xast)} =1-\pout$ regarding $R(\xast)$, we have
\begin{equation}
    R({\bm x}_\ast) = \log_2\left(1+10^{\frac{1}{10}\Gamma_\mathrm{EG}(\xast)}\right),
    \label{eq:outagerate_eg}
\end{equation}
where
\begin{align}
    \Gamma_\mathrm{EG}(\xast) &\triangleq \mathrm{OutageSNR}\left(P_\mathrm{EG}(\xast), \sigma_\mathrm{EG}\right)\label{eq:func_outagesnr}\\
    = P_\mathrm{EG}(\xast) &- N_0 + \sqrt{2}\sigma_\mathrm{EG}(\xast)\erfinv{2\pout-1},
\end{align}
where $\erfinv{\cdot}$ is the inverse error function.
\par
This computation can maximize the rate under the constraint in Eq.\,\eqref{eq:constraint_outage} when $\sigma_x = 0$.
However, its outage probability may exceed the target value under $\sigma_x > 0$ since it does not take the uncertainty in $\tilde{\bm x}_i$ into account the rate prediction.
\section{Rate Selection with Noisy Location Information via NIGP}
\label{sect:proposed}
To consider the location uncertainty for the rate selection problem, we introduce NIGP.
NIGP treats the input uncertainty as an additional output uncertainty based on Taylor expansion for the function of interest\cite{NIPS2011_a8e864d0}.
This section first presents the proposed method based on the original (i.e., first-order approximated) NIGP.
Then, to further accurate rate selection, we extend it to the second-order approximation.
\subsection{Original NIGP-Based Design}
This method first approximates $f(\tilde{\bm x})$ based on a Taylor expansion; i.e.,
\begin{align}
    f(\tilde{\bm x}) &= f({\bm x}) +{\bm \epsilon}^\transpose_x \frac{\partial f({\bm x})}{\partial {\bm x}} + \frac{1}{2}{\bm \epsilon}_x^\transpose \frac{\partial^2 f({\bm x})}{\partial {\bm x}^2} {\bm \epsilon}_x + \cdots \label{eq:nigp_exact}\\
    &\approx f(\tilde{\bm x}) + {\bm \epsilon}_x^\transpose {\bm \partial}_{\overline{f}}(\tilde{\bm x}) + \frac{1}{2}{\bm \epsilon}_x^\transpose {\bm \partial}^2_{\overline{f}}(\tilde{\bm x}) {\bm \epsilon}_x + \cdots, \label{eq:nigp_approx}
\end{align}
where ${\bm \partial}_{\overline{f}}(\tilde{\bm x})$ is a two-dimensional vector.
This vector contains the derivative of the mean of the GP function,
\begin{equation}
    {\bm \partial}_{\overline{f}}(\tilde{\bm x}) = \left[\frac{\partial}{\partial x_1}\overline{f}(\tilde{\bm x}), \frac{\partial}{\partial x_2}\overline{f}(\tilde{\bm x})\right]^\transpose,
\end{equation}
where $\overline{f}(\tilde{\bm x})$ is the predictive mean obtained by Eq.\,\eqref{eq:exactgp_mean} assuming $\tilde{\bm x}$ is the test input ${\bm x}_\ast$, and ${\bm \partial}^2_{\overline{f}}(\tilde{\bm x})$ is its Hessian matrix.
Further, Eq.\,\eqref{eq:nigp_approx} approximates ${\bm x}$ in Eq.\,\eqref{eq:nigp_exact} as $\tilde{\bm x}$ since it cannot access to the latent variable ${\bm x}$.
As can be seen from the second and subsequent terms in Eq.\,\eqref{eq:nigp_approx}, the NIGP regards effects of the input noise, ${\bm \epsilon}_x$, as the output noise appended to $f(\tilde{\bm x})$ .
Based on its first-order approximation, the original NIGP models the observation $y$ as
\begin{align}
    y(\tilde{\bm x}) &\approx f(\tilde{\bm x}) + {\bm \epsilon}_x^\transpose {\bm \partial}_{\overline{f}}(\tilde{\bm x}) + \epsilon_y,
\end{align}
The probability of an observation $y$ is therefore,
\begin{equation}
    p(y\mid f) \approx \normal{f}{\sigma^2_y + {\bm \partial}_{\overline{f}}(\tilde{\bm x})^\transpose \Sigma_x {\bm \partial}_{\overline{f}}(\tilde{\bm x})},
\end{equation}
where $\Sigma_x = \diag{\sigma_{x}, \sigma_{x}}$.
Finally, its mean and variance can be derived as,
\begin{align}
    P_\mathrm{NG1}({\bm x}_\ast) &\triangleq \mathbb{E}\left[\Prx({\bm x}_\ast)\mid \mathcal{D}\right] \approx \mathrm{GP_{mean}}\left({\bf K}'_{\mathrm{NG1}}\right),\label{eq:nigp1_mean}\\
    \sigma^2_\mathrm{NG1}({\bm x}_\ast) &\triangleq \mathrm{Var}\left[\Prx({\bm x}_\ast)\mid \mathcal{D}\right]
    \approx \mathrm{GP_{var}}\left({\bf K}'_{\mathrm{NG1}}\right),\label{eq:nigp1_var}
\end{align}
where
\begin{equation}
    {\bf K}'_{\mathrm{NG1}}= {\bf K}' + \diag{{\bm \partial}^\transpose_{\overline{f}}(\tilde{\bm x}_i) \Sigma_x {\bm \partial}_{\overline{f}}(\tilde{\bm x}_i) \,\Bigg|\, i \in \mathcal{S}},
\end{equation}
and $\mathcal{S} = \{1, 2, \cdots, N\}$.
As with the exact GP, the maximum transmission rate can be estimated from Eq.\,\eqref{eq:outagerate_eg} by replacing $\Gamma_\mathrm{EG}({\bm x}_\ast)$ with $\Gamma_\mathrm{NG1}({\bm x}_\ast)$; i.e.,
\begin{align}
    \Gamma_\mathrm{NG1}(\xast) =
    \mathrm{OutageSNR}\left(P_\mathrm{NG1}({\bm x}_\ast), \sigma_\mathrm{NG1}({\bm x}_\ast)\right).
\end{align}

\subsection{Extension to Second-Order Approximation}
Although the original NIGP can treat the location uncertainty with tractability, it is based on several approximations in addition to the Taylor approximation, as can be seen from Eq.\,\eqref{eq:nigp_approx}.
For further reliable rate selection, we extend it to the second-order approximation.
Under the second-order approximation, based on Eq.\,\eqref{eq:nigp_approx}, the observation result at $\tilde{\bm x}$ can be approximated as,
\begin{equation}
    y(\tilde{\bm x}) \approx f(\tilde{\bm x})
    + {\bm \epsilon}_x^\transpose {\bm \partial}_{\overline{f}}(\tilde{\bm x})
    + \frac{1}{2}{\bm \epsilon}_x^\transpose {\bm \partial}^2_{\overline{f}}(\tilde{\bm x}) {\bm \epsilon}_x
    + \epsilon_y.
\end{equation}
Then, its probability density function under a given function $f$ can be given by, 
\begin{align}
    p(y\mid f) \approx &\mathcal{N}\Big(f + \frac{1}{2}\trace{{\bm \partial}^2_{\overline{f}}(\tilde{\bm x})\Sigma_x}, \sigma^2_y + \nonumber\\
    &{\bm \partial}_{\overline{f}}(\tilde{\bm x})^\transpose \Sigma_x {\bm \partial}_{\overline{f}}(\tilde{\bm x}) + \frac{1}{2}\trace{{\bm \partial}^2_{\overline{f}}(\tilde{\bm x})\Sigma_x}^2\Big),
    \label{eq:prob_ng2}
\end{align}
where $\trace{\cdot}$ is the trace operation.
Finally, based on Eq.\,\eqref{eq:prob_ng2}, the predictive mean and variance are given by,
\begin{align}
    P_\mathrm{NG2}({\bm x}_\ast) &\triangleq \mathbb{E}\left[\Prx({\bm x}_\ast)\mid \mathcal{D}\right] \approx \mathrm{GP_{mean}}\left({\bf K}'_{\mathrm{NG2}}\right),\label{eq:nigp2_mean}\\
    \sigma^2_\mathrm{NG2}({\bm x}_\ast) &\triangleq \mathrm{Var}\left[\Prx({\bm x}_\ast)\mid \mathcal{D}\right] \approx \mathrm{GP_{var}}\left({\bf K}'_{\mathrm{NG2}}\right),
\end{align}
where
\begin{align}
    {\bf K}'_\mathrm{NG2} = {\bf K}'_\mathrm{NG1} + \diag{\frac{1}{2}\trace{\left[{\bm \partial}^2_{\overline{f}}({\bm x}_i)\Sigma_x\right]^2} \,\Bigg |\, i\in \mathcal{S}}.
\end{align}
Note that the detailed computation for the Hessian matrix is presented in Appendix.
Finally, as with the original NIGP-based design, we can estimate the maximum from Eq.\,\eqref{eq:outagerate_eg} by replacing $\Gamma_\mathrm{EG}({\bm x}_\ast)$ with $\Gamma_\mathrm{NG2}({\bm x}_\ast)$ where
\begin{align}
    \Gamma_\mathrm{NG2}(\xast) = 
    \mathrm{OutageSNR}\left(P_\mathrm{NG2}({\bm x}_\ast), \sigma_\mathrm{NG2}({\bm x}_\ast)\right).
\end{align}
\section{Numerical Results}
\label{sect:performance}

\begin{table}[t]
    \caption{Simulation parameters.}
    \label{table:simulation_parameters}
    \centering
     \begin{tabular}{p{4.0cm}|p{3cm}}
      \hline
       Parameter & Setting \\\hline
       Number of sensing data $N$ & 100\\
       Number of test points & 20\\
       AWGN $N_0$ & -174\,[dBm/Hz]\\
       Correlation distance $d_\mathrm{cor}$ & 50\,[m]\\
       Shadowing standard deviation $\sigma_\mathrm{dB}$ & 6.0\,[dB]\\
       Path loss index $\eta$ & 3.0\\
       Transmission power $P_\mathrm{Tx}$ & 10\,[dBm]\\
       Target outage probability $\pout$ & $1.0\times 10^{-3}$\\
      \hline
     \end{tabular}
   \end{table}

\begin{figure}[t]
  \centering
    \subfigure[Pure GP.]{
      \includegraphics[width=0.45\linewidth]{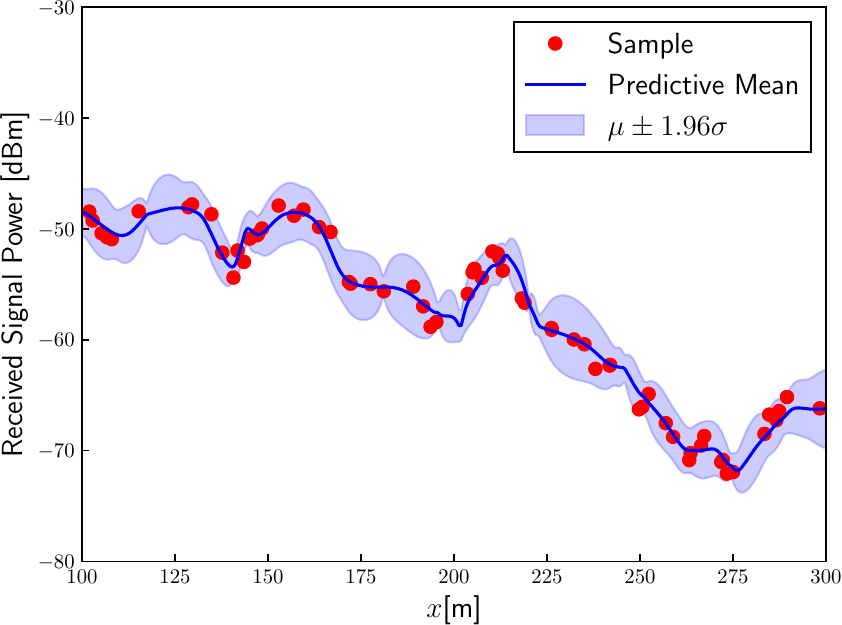}
    \label{subfig:example_puregp}
    }
    \subfigure[Path loss.]{
      \includegraphics[width=0.45\linewidth]{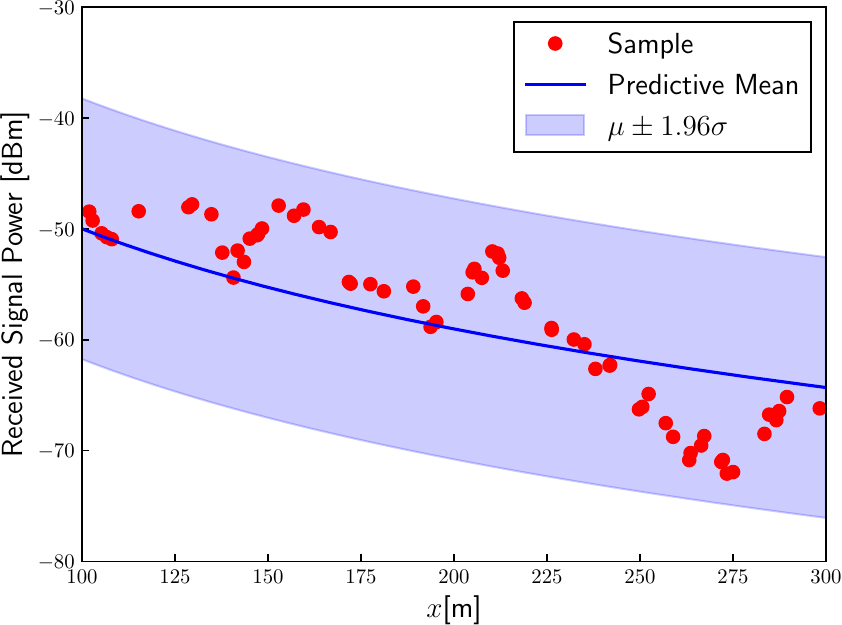}
    \label{subfig:example_pathloss}
    }
    \subfigure[NIGP (1st order).]{
      \includegraphics[width=0.45\linewidth]{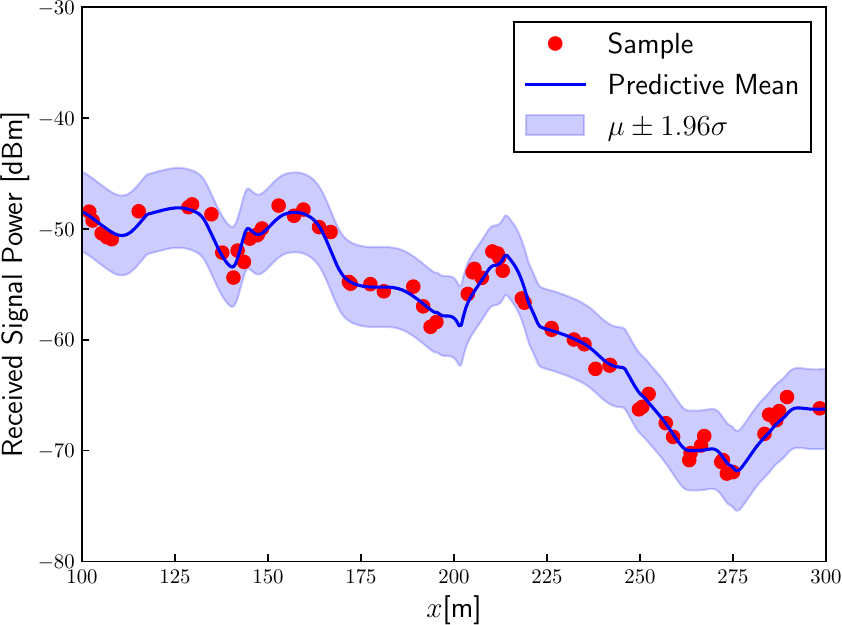}
      \label{subfig:example_nigp1}
    }
    \subfigure[NIGP (2nd order).]{
      \includegraphics[width=0.45\linewidth]{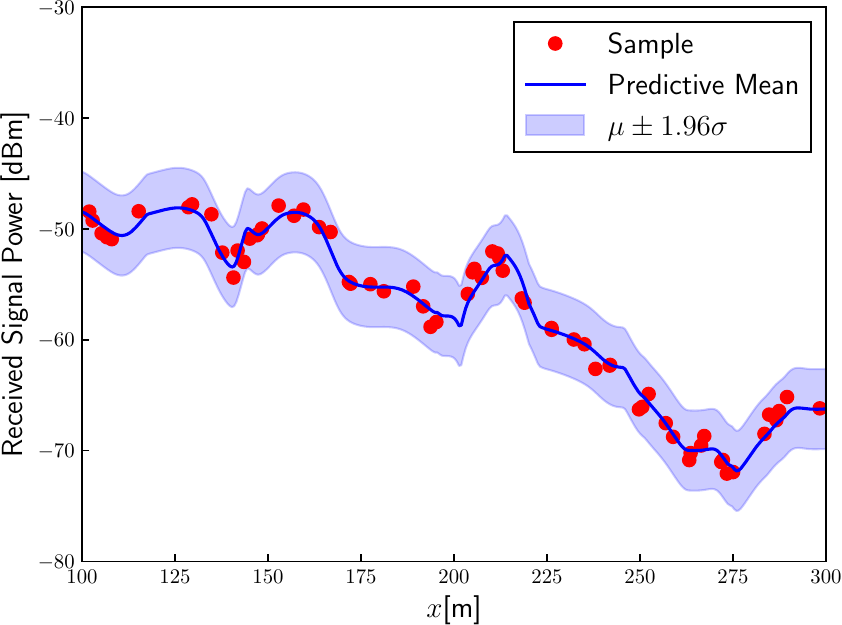}
      \label{subfig:example_nigp2}
    }
  \caption{Prediction examples where $\sigma_x = 10$\,(one-dimensional case).\label{fig:example_prediction}}
\end{figure}

This section presents numerical results for the rate selection methods.
Specifically, we focus on the robustness of the NIGP-based design in terms of outage performance.
Our simulation considers a 300-meter-square area, where a transmitter is deployed at ${\bm x}_\mathrm{Tx} = [-10\,\text{[m]}, 150\,\text{[m]}]$.
The simulation was performed with $10^5$ independent trials; each trial includes channel generation, sensing, rate selection, and outage evaluation.
Further, to discuss the impact of positioning errors in smartphone-based positioning\cite{Zangenehnejad2021}, we set $\sigma_x$ between zero to 20.
Other simulation parameters are summarized in Table~\ref{table:simulation_parameters}.
\par
Fig.\,\ref{fig:example_prediction} summarizes examples of radio map construction for the case where $\sigma_x = 10$.
These figures illustrate the predictive mean and variance values for the estimated received signal power.
The colored region represents $\mu({\bm x}) \pm 1.96\sigma({\bm x})$, where $\mu({\bm x})$ and $\sigma({\bm x})$ denote the predictive mean and standard deviation, respectively.
These examples depict one-dimensional radio map construction results, assuming all receivers perform sensing along the line $x_2 = 150$, for demonstration purposes.
Based on Eqs.~\eqref{eq:exactgp_mean}\eqref{eq:nigp1_mean}, pure GP (Fig.~\ref{subfig:example_puregp}) and NIGP with 1st-order approximation (Fig.~\ref{subfig:example_nigp1}) predict identical mean values.
However, pure GP predicts a narrower variance than NIGP, indicating that it tends to be overly confident in its regression results.
Furthermore, similar to the NIGP with 1st-order approximation, the NIGP with 2nd-order approximation (Fig.~\ref{subfig:example_nigp2}) also predicts a wider variance for the received signal power than pure GP.
Note that the path-loss-based method (Fig.~\ref{subfig:example_pathloss}) sets the shadowing standard deviation $\sigma_{\mathrm{dB}}$ as its prediction uncertainty, as it does not depend on the dataset $\mathcal{D}$.
Although this method can satisfy the outage constraint, its pessimistic prediction can degrade rate performance.
\par
\begin{figure}
    \centering
    \includegraphics[width=1.0\linewidth]{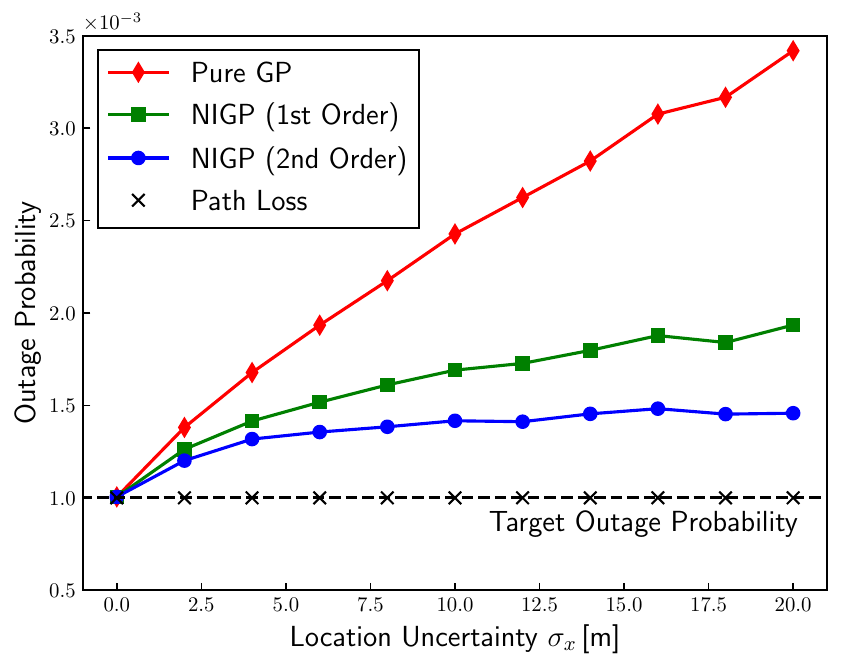}
    \caption{Effects of location uncertainty on outage performance.}
    \label{fig:effect_sigmax}
\end{figure}
Effects of location uncertainty on the outage probability performance are shown in Fig.\,\ref{fig:effect_sigmax}.
We changed the standard deviation of location noise $\sigma_x$ from zero to 20.
All methods, including pure GP, can satisfy the outage constraint at $\sigma_x=0$.
However, increasing $\sigma_x$ exceeds the outage performance of GP-based methods.
This degradation effect was significant in the pure GP; for example, $2.5\times 10^{-3}$ at $\sigma_x=10$ and $3.5\times 10^{-3} $ at $\sigma_x=20$.
Although NIGP-based methods are similarly affected by location uncertainty regarding outage performance, their performance degradation remains robust against the intensity of $\sigma_x$.
In particular, by extending NIGP to the 2nd-order approximation, the gap to the target outage probability can be suppressed to approximately $0.4 \times 10^{-3}$.
The 2nd-order NIGP enables rate predictions almost satisfying the desired outage probability, regardless of location uncertainty.
\par
\begin{figure}[t]
  \centering
    \subfigure[$\sigma_x=10\,\text{[m]}$.]{
      \includegraphics[width=0.46\linewidth]{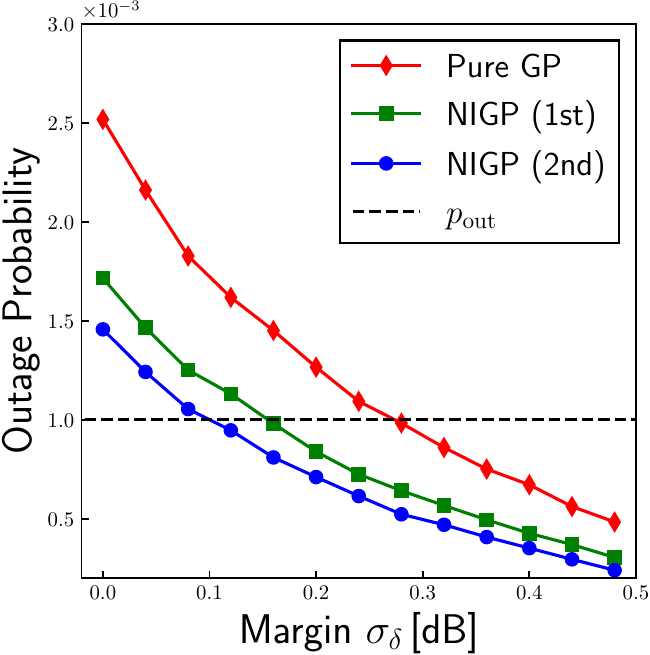}
    \label{subfig:effect_margin_sigmax10}
    }
    \subfigure[$\sigma_x=20\,\text{[m]}$.]{
      \includegraphics[width=0.46\linewidth]{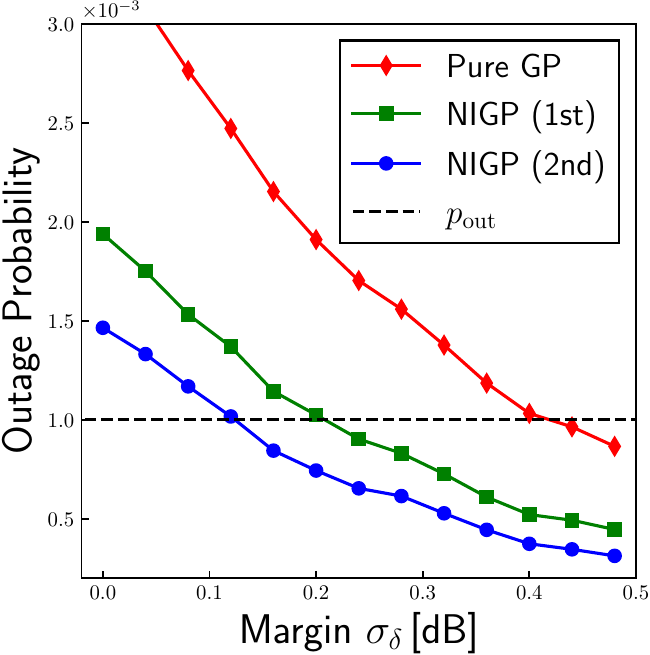}
    \label{subfig:effect_margin_sigmax20}
    }
  \caption{Effects of $\sigma_\delta$ on outage performance. The error in the radio map construction is widely estimated as $\sigma_\delta$ increases, thereby the outage probability decreases.\label{fig:effects_margin}}
\end{figure}
However, since NIGP relies on the Taylor approximation, this method slightly exceeds the target outage probability.
This suggests that some compensation will be necessary to achieve the outage constraint strictly.
Thus, we introduce an additional margin to the predictive standard deviation to address this.
Let us define a hyperparameter $\sigma_{\delta} (\geq 0)$, which is uniformly added to the predictive standard deviation.
This approach results in a more conservative rate prediction, improving outage probability performance.
Effects of $\sigma_{\delta}$ on the outage probability performance are shown in Fig.\,\ref{fig:effects_margin}.
To clarify impacts of $\sigma_x$, we evaluated the performance under $\sigma_x=10$ (Fig.\,\ref{subfig:effect_margin_sigmax10}) and $\sigma_x=20$ (Fig.\,\ref{subfig:effect_margin_sigmax20}).
The pure GP exhibited dependence on $\sigma_x$ regarding the required margin to satisfy the outage constraint, showing variations such as $\sigma_\delta \approx 0.44$ for $\sigma_x=20$ and $\sigma_\delta \approx 0.28$ for $\sigma_x=10$.
In contrast, NIGP with the 2nd-order approximation showed low sensitivity to $\sigma_x$, with margins of approximately $\sigma_\delta \approx 0.16$ for $\sigma_x=20$ and $\sigma_\delta \approx 0.12$ for $\sigma_x=10$.
Hence, the NIGP-based design offers an advantage over pure GP regarding reduced sensitivity to location uncertainty, simplifying hyperparameter design in practical scenarios.
\par
\begin{figure}[t]
    \centering
    \includegraphics[width=1.0\linewidth]{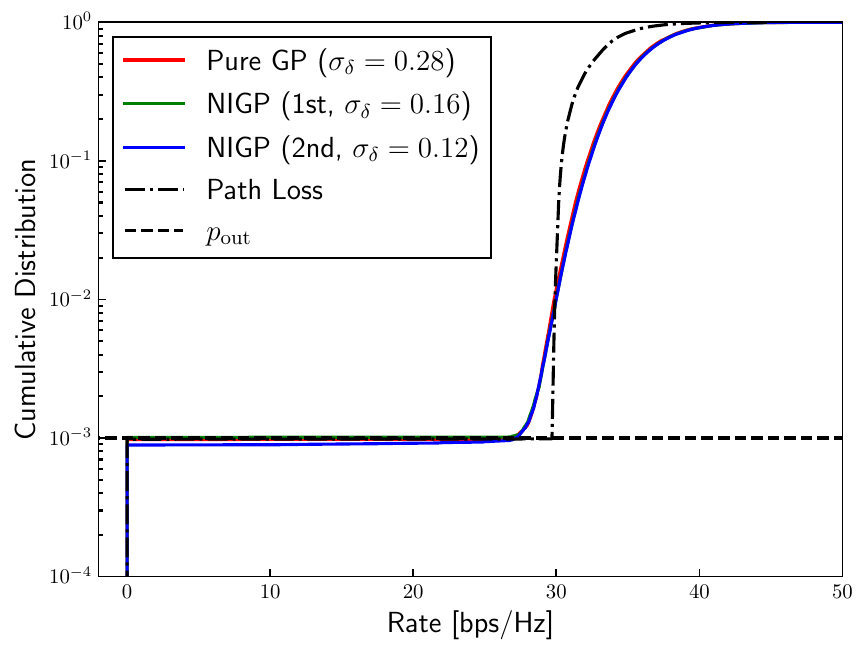}
    \caption{Cumulative distributions of the receiving rate where $\sigma_x=10$. GP-based methods appended $\sigma_\delta$ to their predictive standard deviations so that the outage constraint satisfies $\pout=1.0\times10^{-3}$; the values of $\sigma_\delta$ are determined based on the results in Fig.\,\ref{fig:effects_margin}.}
    \label{fig:cumulative_distribution}
\end{figure}

Finally, we evaluate the receiving rate performance.
As seen from the radio map construction example using the path loss-based method (Fig.\,\ref{subfig:example_pathloss}), it is straightforward to satisfy the outage constraint by conservatively estimating the received signal power.
However, underestimating the received power directly leads to a reduction in transmission rate; hence, it is essential to evaluate not only outage probability but also the statistical characteristics of the achievable rates.
To address this issue, we present the cumulative distribution of received rates at the test points.
The cumulative distribution performance is shown in Fig.\,\ref{fig:cumulative_distribution}.
This figure plots the receiving rates at the test points obtained over $10,000$ trials.
Note that this evaluation assumed the receiving rate to be zero when the predictive rate exceeded the actual channel capacity.
Additionally, to assess the impact of the Taylor approximation used in NIGP, we introduced $\sigma_\delta$ to the GP-based methods, each tuned to satisfy the target outage constraint. The values of $\sigma_\delta$ for each method were designed based on Fig.\,\ref{fig:effects_margin}.
\par
The results show that although the path loss-based method has an advantage up to a cumulative probability of around $10^{-2}$, GP-based methods outperform it in the region beyond this level, demonstrating improved received rates across most regions.
Moreover, despite the rate predictions of NIGP-based methods relying on the Taylor approximation, their rate performance was found to be comparable to pure GP when $\sigma_\delta$ was appropriately selected to satisfy the outage constraint.
\par
In summary, the NIGP-based method can approximately achieve the target outage probability regardless of the intensity of location uncertainty. Moreover, the gap between the actual and target outage probabilities can be reduced by increasing the order of the Taylor approximation.
Furthermore, NIGP showed low sensitivity to location uncertainty in the additional margin $\sigma_\delta$ required for achieving the target outage probability strictly, simplifying hyperparameter design compared to pure GP.
Additionally, the statistical properties of the receiving rates with NIGP are comparable to those of pure GP under appropriately chosen $\sigma_\delta$.
Thus, employing NIGP simplifies and enhances the efficiency of radio map utilization under conditions of location uncertainty.

\section{Conclusion}
This paper proposed the NIGP-based rate selection.
Unlike a pure GP-based design, it can enhance robustness against localization errors in rate selection problems with outage constraints. We also showed that extending NIGP to the second-order approximation can further mitigate the impact of location uncertainty without significant rate degradation.
NIGP will simplify the pre-design of radio map-based high-reliability communications by reducing dependency on the strength of location uncertainty.

\appendix
\label{sect:appendix}
\subsection{Hessian Matrix Computation}
The Hessian matrix at the coordinate ${\bm x}=[x_1, x_2]^\transpose$ is
\begin{equation}
{\bm \partial}^2_{\overline{f}}({\bm x})=
\begin{pmatrix}
\displaystyle \frac{\partial^2 \overline{f}({\bm x})}{\partial x_1^2} & \displaystyle \frac{\partial^2 \overline{f}({\bm x})}{\partial x_1\partial x_2} \\[1ex]
\displaystyle \frac{\partial^2 \overline{f}({\bm x})}{\partial x_2\partial x_1} & \displaystyle \frac{\partial^2 \overline{f}({\bm x})}{\partial x_2^2}
\end{pmatrix}.
\end{equation}
According to Eq.\,\eqref{eq:exactgp_mean}, the GP posterior mean function can be regarded as a weighted sum for the observation data; i.e.,
\begin{equation}
    \overline{f}({\bm x}) = \overline{P}({\bm x}) + \sum_{i=1}^{N} \alpha_i\, k({\bm x},{\bm x}_i),
\end{equation}
where $[\alpha_1, \alpha_2, \cdots, \alpha_N] = \left({\bf K}'\right)^{-1}\left({\bf Y} - \overline{\bf P}\right).$
Then, the second derivative with respect to $x_m, x_n$ is given by
\begin{equation}
    \frac{\partial^2 \overline{f}({\bm x})}{\partial x_m \partial x_n} = \sum_{i=1}^{N} \alpha_i\, \frac{\partial^2 k({\bm x},{\bm x}_i)}{\partial x_m \partial x_n}.
\end{equation}
Based on Eq.\,\eqref{eq:exponential_kernel}, we design the kernel as $k({\bm x},{\bm x}_i) = \sigma_f^2\,\exp\left(-\frac{r_{c,i}}{l}\right)$,
where $r_{c,i} = \distance{{\bm x}}{{\bm x}_i} + d_c$ and $d_c$ is a positive-valued small constant; note that we add $d_c$ to the distance to avoid discontinuity point at $\distance{{\bm x}}{{\bm x}_i}=0$.
Thus, when ${\bm x}_{i}=[x_{i,1}, x_{i,2}]^\transpose$, the partial differentiation for the kernel function can be given by
\begin{align}
    \frac{\partial^2 k({\bm x},{\bm x}_i)}{\partial x_m \partial x_n} &= \sigma_f^2\,\exp\left(-\frac{r_{c,i}}{l}\right)\times\\
    \Bigg[-\frac{\delta_{mn}}{l r_{c,i}}&+\frac{(x_m-x_{i,m})(x_n-x_{i,n})}{r_{c,i}^2}\left(\frac{1}{l^2}+\frac{1}{l\,r_{c,i}}\right)\Bigg],\nonumber
\end{align}
where $\delta_{mn}$ denotes the Kronecker delta.

\bibliographystyle{IEEEbib}
\bibliography{reference}

\begin{thebibliography}{10}

\bibitem{zeng_comst2024}
Y.~Zeng et~al.,
\newblock ``A tutorial on environment-aware communications via channel
  knowledge map for {6G},''
\newblock {\em IEEE Commun. Surv. Tuts.}, vol. 26, no. 3, pp. 1478--1519, 2024.

\bibitem{zhao-dyspan2007}
Y.~Zhao et~al.,
\newblock ``Applying radio environment maps to cognitive wireless regional area
  networks,''
\newblock in {\em Proc. IEEE DySPAN 2007}, 2007, pp. 115--118.

\bibitem{perez_iotj2025}
D.~E. Pérez, O.~L.~A. López, and H.~Alves,
\newblock ``{EVT}-enriched radio maps for ultrareliable communication,''
\newblock {\em IEEE Internet Things J.}, vol. 12, no. 12, pp. 22012--22022,
  2025.

\bibitem{tian_jiot2025}
J.~Tian, L.~Cong, and H.~Qin,
\newblock ``Simultaneous localization and rough indoor floorplan mapping
  combining {PDR} and {FM/Wi-Fi} radio signals,''
\newblock {\em IEEE Internet Things J.}, vol. 12, no. 13, pp. 24750--24763,
  2025.

\bibitem{wang_twc2024}
J.~Wang et~al.,
\newblock ``Sparse bayesian learning-based hierarchical construction for {3D}
  radio environment maps incorporating channel shadowing,''
\newblock {\em IEEE Trans. Wireless Commun.}, vol. 23, no. 10, pp.
  14560--14574, 2024.

\bibitem{Rasmussen2004}
C.~E. Rasmussen,
\newblock {\em Gaussian Processes in Machine Learning},
\newblock Springer Berlin Heidelberg, Berlin, Heidelberg, 2004.

\bibitem{kallehauge_gcom2022}
T.~Kallehauge et~al.,
\newblock ``Predictive rate selection for ultra-reliable communication using
  statistical radio maps,''
\newblock in {\em Proc. IEEE GLOBECOM2022}, 2022, pp. 4989--4994.

\bibitem{sato-tccn2017}
K.~Sato and T.~Fujii,
\newblock ``Kriging-based interference power constraint: Integrated design of
  the radio environment map and transmission power,''
\newblock {\em IEEE Trans. Cogn. Commun. Netw.}, vol. 3, no. 1, pp. 13--25,
  2017.

\bibitem{5711699}
E.~Dall'Anese, S.-J. Kim, and G.~B. Giannakis,
\newblock ``Channel gain map tracking via distributed {Kriging},''
\newblock {\em IEEE Trans. Veh. Technol.}, vol. 60, no. 3, pp. 1205--1211,
  2011.

\bibitem{Cressie}
N.~A.~C. Cressie,
\newblock {\em Statistics for spatial data},
\newblock Wiley-Interscience, 1993.

\bibitem{Zangenehnejad2021}
F.~Zangenehnejad and Y.~Gao,
\newblock ``{GNSS} smartphones positioning: advances, challenges,
  opportunities, and future perspectives,''
\newblock {\em Satell. Navig.}, vol. 2, pp. 24, 2021.

\bibitem{NIPS2011_a8e864d0}
A.~Mchutchon and C.~Rasmussen,
\newblock ``Gaussian process training with input noise,''
\newblock in {\em Advances in Neural Information Processing Systems},
  J.~Shawe-Taylor, R.~Zemel, P.~Bartlett, F.~Pereira, and K.Q. Weinberger, Eds.
  2011, vol.~24, Curran Associates, Inc.

\bibitem{yi-tnse2024}
S.~Yi, H.~Zhang, and K.~Liu,
\newblock ``{V2IViewer}: Towards efficient collaborative perception via point
  cloud data fusion and vehicle-to-infrastructure communications,''
\newblock {\em IEEE Trans. Netw. Sci. Eng.}, vol. 11, no. 6, pp. 6219--6230,
  2024.

\bibitem{Gudmundson-el1991}
M.~Gudmundson,
\newblock ``Correlation model for shadow fading in mobile radio systems,''
\newblock {\em Electron. Lett.}, vol. 27, pp. 2145--2146, Nov. 1991.

\end{thebibliography}

\end{document}